# Modeling of anti-tumor immune response: immunocorrective effect of weak centimeter electromagnetic waves

O.G. Isaeva[a,*] and V.A. Osipov[a]

[a] *Bogoliubov Laboratory of Theoretical Physics, Joint Institute for Nuclear Research, 141980, Dubna, Moscow Region, Russia*

**Abstract**

We formulate the dynamical model for the anti-tumor immune response based on intercellular cytokine-mediated interactions with the interleukin-2 (IL-2) taken into account. The analysis shows that the expression level of tumor antigens on antigen presenting cells has a distinct influence on the tumor dynamics. At low antigen presentation a progressive tumor growth takes place to the highest possible value. At high antigen presentation there is a decrease in tumor size to some value when the dynamical equilibrium between the tumor and the immune system is reached. In the case of the medium antigen presentation both these regimes can be realized depending on the initial tumor size and the condition of the immune system. A pronounced immunomodulating effect (the suppression of tumor growth and the normalization of IL-2 concentration) is established by considering the influence of low-intensity electromagnetic microwaves as a parametric perturbation of the dynamical system. This finding is in qualitative agreement with the recent experimental results on immunocorrective effects of centimeter electromagnetic waves in tumor-bearing mice.

---

[*] Corresponding author. Current address: Bogoliubov Laboratory of Theoretical Physics, Joint Institute for Nuclear Research, 141980, Dubna, Moscow Region, Russia. Fax: (7)(49621) 65084. E-mail addresses: issaeva@theor.jinr.ru, osipov@theor.jinr.ru

## 1. Introduction

A theoretical investigation of cancer growth under immunological activity has a long history (see, e.g., [1] and the references therein). Most of the known models consider dynamics of two main populations: effector cells and tumor cells [27,44]. Some models include the dynamics of certain cytokines [3,10,24]. An important issue of these studies is a variation of the concentration of cytokines during the disease. As is known, tumor growth results in imbalance between the production and the regulation of cytokines as well as in the reduction of the corresponding receptors thus leading to the suppression of the immunological activity. Therefore, the methods for enhancement of both the anti-tumor resistance and the general condition of the immune system are of current clinical and theoretical interest. One of them refers to the use of cytokines, in particular interleukin-2 (IL-2) [16,18,21,22]. Interleukin-2 is considered as the main cytokine responsible for the proliferation of cells containing IL-2 receptors and their following differentiation [48]. IL-2 is mainly produced by activated CD4+ T-cells. There are many evidences that IL-2 plays an important role in specific immunological reactions to alien agents including tumor cells [28,38,48]. Clinical trials also show positive treatment effects at low doses of IL-2 [18, 40—42]. At the same time, at high doses treatment with IL-2 may cause serious hematologic violations revealed by anemia, granulocytopenia, thrombocytopenia, and lymphocytosis.

The first detailed model of the anti-tumor immune response with IL-2 taken into account was proposed by DeBoer et al. [10]. It contains eleven ordinary differential equations and five algebraic equations and was used to study the role of macrophage- T lymphocyte interactions that are involved in the cellular immune response. The analysis shows a possibility for both tumor regression and uncontrolled tumor growth depending on "the degree of antigenicity" (the initial size of the T lymphocyte precursor populations that can be stimulated upon introduction of specific antigen).

Afterward, Kirschner and Panetta proposed a simpler model where only three main populations were considered: the effector cells, the tumor cells, and IL-2 [24]. The model allows them to study effects of immunotherapy based on the use of cytokines together with adoptive cellular immunotherapy (ACI). ACI refers to the injection of cultured immune cells that have anti-tumor reactivity into tumor bearing host [24]. It was found that without immunotherapy the immune system is unable to clear the tumor with low antigenicity (a measure of how different the tumor is from 'self'), while for highly antigenic tumors reduction to a small dormant tumor takes place. When tumor exhibits average antigenicity, stable limit cycles were observed. This implies that the tumor and the immune system undergo oscillations.

Further, in the framework of the model by Kirschner and Panetta [24], Arciero et al. considered a novel treatment strategy known as small interfering RNA (siRNA) therapy [3]. The model [3] consists of a system of nonlinear, ordinary differential equations describing tumor cells, immune effectors, the immuno-stimulatory and suppressive cytokines IL-2 and TGF-β as well as siRNA. TGF-β suppresses the immune system by inhibiting the activation of effector cells and reducing tumor antigen receptors. It also stimulates tumor growth by promoting angiogenesis. siRNA treatment suppresses TGF-β production by targeting the mRNA that codes for TGF-β, thereby reducing the presence and effect of TGF-β in tumor cells. The model predicts conditions under which siRNA treatment can be successful in returning TGF-β producing tumors to its passive, non-immune evading state.

Recently, a recovery of IL-2 production after the exposure of tumor-bearing mice to low-intensity centimeter waves was experimentally observed [17]. This indicates that exposure to centimeter electromagnetic waves may be used for an enhancement of the anti-tumor immune response. In experiments, solid tumors were formed by means of hypodermic transplantation of the ascitic Ehrlich's carcinoma cells. Notice that previous investigations of effects of low-intensive microwave radiation also show the immunomodulating effects at certain frequency ranges and intensities (see, e.g., [9,26]). These findings stimulate our interest to study the influence of weak centimeter electromagnetic waves on tumor-immune dynamics. Actually, the influence of EMR depends on the type of radiation, a distance from the radiation source (far-field versus near-field exposure conditions), frequency range, sizes and shapes of objects. Evidently, it is a hard problem to take properly into account all these factors within any theoretical description. In this paper, we offer a reasonable phenomenological approach.

First of all, we formulate an appropriate mathematical model of anti-tumor immune response with the interleukin-2 taken into account (Sect. 2). To this end, we follow the scheme of intercellular cytokine mediated interaction in cellular immune response proposed by Wagner et al. in [48] which was modified by taking into account co-stimulatory factors such as B7/CD28 and CD40/CD40L instead of interleukin-1 (see, e.g., [28,38]). The analysis of the model is presented in Section 3. In Section 4 we discuss a possibility of immunomodulating effect of weak radiofrequency electromagnetic radiation (RF EMR) considering the influence of irradiation as a parametric perturbation of the initial dynamical system.

## 2 Model

We describe the dynamics of cellular populations participating in formation of cytotoxic effector cells and cytokines mediating these reactions in accordance with a scheme presented in Figure1. Some important remarks should be done. Generally, the population of T cells is divided into two subpopulations: helper T cells (HTL) that express marker CD4 on their surface and cytotoxic T cells (CTL) that express CD8 marker [38]. CTL specifically recognize complexes of antigen (AG) with MHC (major histocompatibility complex, in human being — HLA human lymphocyte antigens) class I on the surface of alien or tumor cell and destroy them through this interaction. In contrast to CTL, HTL recognize complexes AG-MHC II on tumor cell and play a regulatory role in the expansion of CTL.

In order to stimulate both HTL and CTL against tumor antigen, it must be presented via MHC class I and II molecules expressed by professional antigen-presenting cell (APC). There are three main types of professional antigen-presenting cells: dendritic cells, macrophages and B cells. Dendritic cells and, to a lesser extent, macrophages have the broadest range of antigen presentation and are probably the most important APC. They exist as immature (iAPC) and mature (mAPC) forms.

The dynamical equations for immature APC ($m$) and mature APC ($M$) are written as

$$\dot{m} = V_m - \beta_m m - \gamma_m mT, \tag{1}$$

$$\dot{M} = \gamma_m mT - \beta_M M. \tag{2}$$

In (1) $V_m$ characterizes a steady inflow of iAPC from monocytes which in turn are formed from stem cells in the bone marrow. The second term describes iAPC death rate. iAPC phagocytose AG, degrade it, and present their fragments at the plasma membrane using MHC molecules upon maturation. Simultaneously, they express co-stimulatory molecules such as B7 and CD40 [28,38]. Thus iAPC become mature APC (mAPC) expressing both complexes AG-MHC-I and II as well as co-stimulatory molecules which are recognized by specific receptors on T cells. The rate of transfer from iAPC to mAPC is described by the third term in (1) where $T$ is the number of tumor cells. The concentration of antigen is supposed to be proportional to the number of tumor cells. The production rate of mAPC in (2) is equal to the rate of transfer. The mAPC death rate is described by the second term.

Dynamics of HTL precursors ($H$) and IL-2 ($I_2$) is chosen to be

$$\dot{H} = V_H - \beta_H H, \qquad (3)$$
$$\dot{I}_2 = \gamma_H HM - \tilde{\alpha}_L L I_2 - \gamma_T T I_2. \qquad (4)$$

In (3) $V_H$ characterizes the inflow of HTL precursors (HTLP) from stem cells. The second term shows the death rate of HTLP. As a result of interaction between complex AG-MHC-II on mAPC and HTLP (see signal 1 in Figure1) in the presence of a number of co-stimulatory molecules CD40 (signal 2), activated HTLP produce lymphokines (including IL-2) and corresponding receptors. A similar production is observed when mAPC presents antigen with MHC-I molecule to cytotoxic T cells precursor (CTLP) in the presence of co-stimulatory molecules B7 binding to CD28 markers. The interaction between IL-2 and corresponding receptors on activated T lymphocyte precursors (HTLP and CTLP) induces their proliferation and differentiation into mature T lymphocytes (HTL and CTL). In order to simplify consideration we omit an equation for HTL activated by tumor antigen assuming that the proliferation of HTL in response solely to IL-2 is absent. This is based on the fact that the levels of expression of IL-2 receptors on HTL are substantially lower than those observed on CTL [30]. Hence, the new AG stimulation is required to support the proliferation of HTL. In addition, our analysis shows that the exclusion of the equation for activated HTL does not influence the tumor growth dynamics within the model. Thus, HTLP stimulated by mAPC are assumed to perform the role of IL-2 producers. We suggest that the concentration of IL-2 grows linearly with HTLP and mAPC (first term in (4)). As long as IL-2 is a short-distance cytokine, it is supposed that target cells CTL ($L$)) effectively consume IL-2. For this reason, we neglect the term presenting loss rate of IL-2. We also consider in (4) the diminution of IL-2 molecules (forth term) as a result of interaction with prostaglandins, immuno-suppressing substances which both suppress the production of IL-2 and directly destroy its molecules [37]. Notice that concentration of prostaglandins is supposed to be proportional to the number of tumor cells $T$.

Let us formulate the dynamical equation for cytotoxic T cells ($L$). Similarly to [7,13,27,31,32,34] we suggest that CTL-tumor cell interaction follows enzymatic kinetics, that is

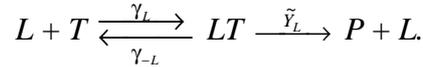

$$L + T \underset{\gamma_{-L}}{\overset{\gamma_L}{\rightleftarrows}} LT \xrightarrow{\tilde{\gamma}_L} P + L.$$

Indeed, CTL can be bound to tumor cell either reversibly (forward and back reactions with the corresponding rates $\gamma_L LT$ and $\gamma_{-L}(LT)$, tumor cells are not 'suffering') or irreversibly ($LT$ complex is formed) inducing cell death [2,38]. CTL kill tumor cells via one of two main mechanisms. The first one is based on the secretion of perforines. Perforines are embedded into the membrane of tumor cells and form pores thus clearing a way for penetrating water. $LT$ complex dissociates into 'doomed' tumor cell ($P$) and CTL ($L$) with a rate $\tilde{\gamma}_L (LT)$. Tumor cell swells and gets killed while CTL looks for the new target. The second mechanism involves programmed cell death (apoptosis) through the Fas/Fas ligand pathway. Thus, we introduce an additional equation for "substratum-enzyme complex" ($LT$) and the equation for immune cells reads

$$\dot{L} = V_L - \beta_L L + \alpha_L L I_2 - \gamma_L LT + \gamma_{-L}(LT) + \tilde{\gamma}_L(LT) \qquad (5)$$

$$(\dot{LT}) = \gamma_L LT - \gamma_{-L}(LT) - \tilde{\gamma}_L(LT). \qquad (6)$$

In (5) $V_L$ characterizes the constant inflow of CTL into the tissue. The second term describes the death rate of CTL. The population of CTL increases due to its proliferation in the presence of IL-2 (the third term in (5)). The remaining terms in (5) describe the CTL-tumor cell interactions. As is seen, equation (5) describes an expansion of cytotoxic T cells in the presence of IL-2 without antigen stimulation. A similar consideration was presented within the programmed proliferation model by Wodarz and Thomsen in [50]. They suggested that the interaction with infected cell transfer CTLP to population of proliferating CTL, which undergo a limited number of divisions without AG stimulation before the differentiation into effector cells. Thus, they use separate equations for population of CTLP, effector cells and $n$ intermediate populations of CTL that passed $i = 1, 2, \dots n$ divisions.

Finally, the population of tumor cells is described by

$$\dot{T} = -\alpha_T T \ln \frac{\beta_T T}{\alpha_T} - \gamma_L TL + \gamma_{-L}(LT). \tag{7}$$

Notice that some studies include exponential law to describe tumor growth (see, e.g., [13,43,44]). When tumor cells grow in conditions of an interior competition one has to use the limiting growth laws, for instance logistic or Gompertzian [24,27,46]. In our model, we prefer to choose the Gompertzian law (the first term in (7)). This allows us to avoid the regime of tumor autoregression under immunological activity only. Such outcome would contradict numerous clinical experiments. As another reason, clinical and experimental observations show that the growth of some tumors is fitted by the Gompertzian function [19,33].

It should be mentioned that we do not consider here processes of angiogenesis (vascular growth), invasion and metastasis, which are of importance at late (III-IV) stages of the tumor growth. Actually, inclusion of processes of vascular growth and invasion requires serious extension of the model to describe dynamics of cytokines, enzymes and other components regulating these processes. Besides, it would be necessary to take into account spatial migration of cell populations during the process of invasion (see, e.g., [8]). Therefore, the system (1)—(7) is valid for the description of early stages of the tumor growth when the processes of angiogenesis, invasion and metastasis are not of critical importance. This model allows us to study the different regimes of early immunological activity. However, the formulated model consists of seven differential equations and a great number of model parameters. This makes it difficult to analyze even qualitatively. Therefore, trying to decrease the number of the model equations we will make some simplifying assumptions.

First of all, for lingering diseases one can consider $(LT)(t)$ as rapid variable. In other words, it rapidly reaches its stationary value which remains fixed during the time of the immune response. In this case, $\overline{(\dot{LT})} = 0$ and one obtains from (6) that $\overline{(LT)} = \gamma_L LT / (\tilde{\gamma}_L + \gamma_{-L})$. Substituting this expression in (5) and (7) one finally gets

$$\dot{L} = V_L + \alpha_L L I_2 - \beta_L L, \tag{8}$$

$$\dot{T} = -\alpha_T T \ln \frac{\beta_T T}{\alpha_T} - \gamma'_L LT, \tag{9}$$

where $\gamma'_L = \gamma_L \tilde{\gamma}_L / (\gamma_{-L} + \tilde{\gamma}_L)$.

Let us assume that $m(t)$, $M(t)$, and $H(t)$ are also in quasi steady states. In this case, equation (4) is written as

$$\dot{I}_2 = \frac{\alpha_{I_2} T}{T + K_T} - \tilde{\alpha}_L L I_2 - \gamma_T T I_2, \tag{10}$$

where expressions for $\alpha_{I_2} = \gamma_H V_H V_m / (\beta_H \beta_M)$ and $K_T = \beta_m / \gamma_m$ follow from the equations $\dot{m} = 0$, $\dot{M} = 0$, and $\dot{H} = 0$.

Finally, the model becomes much simpler and contains only three main equations (8)—(10). Nevertheless, it incorporates the most important modern concepts of tumor-immune dynamics including the influence of IL-2 dynamics. The first two equations resemble the famous predator-pray model with tumor cells as 'victims'. As is seen, the growth rate of 'predators' (CTL population) depends on the concentration of IL-2. In (8) we take into account the steady influx of CTL likewise some other considerations (see, e.g., [27,44]). Let us mention once more that, at first glance, such description ignores the preliminary antigen stimulation. In fact, this stimulation is considered in (10) through the first term where IL-2 production depends on tumor size. We use the hyperbola which allows us to take into account a limitation in stimulation of the immune system by the growing tumor. At small $T$ the growth rate is linear in tumor size while for big tumor ($T \gg K_T$) it tends to a constant value. The last term in (10) reflects a destruction of IL-2 by metabolic products of tumor cell which are proportional to the concentration of tumor cells [37].

*2.1. Parameter set*

An important question is the choice of parameters. The dynamics of disease is very sensitive to parameters in equations (8)—(10). The used values are given in Table 1. Some values were estimated by using the available experimental data. In particular, the growth parameters of ascitic Ehrlich carcinoma $\alpha_T$ and $\beta_T$ were obtained from the experimental data found in Lobo's results where the Ehrlich ascites tumor cell line was cultured *in vitro* [29]. Using the least-squares method, we fitted the experimental data by Gompertzian curve. The death rate of CTL was estimated using the relation $\beta_L = 1/\tau$ where $\tau$ is their known average lifetime. The rate of steady inflow of CTL was calculated from the relation $V_L = \beta_L L_{\text{free}}$ where $L_{\text{free}}$ (the number of CTL capable to recognize carcinoma specific antigen in the organism without tumor) was estimated to be about $2.4 \times 10^5$ cells using the data for the number of CD8+ T cells in spleen of mice [5] and a percent value of T cells specific for tumor type [14]. For the rest of parameters we chose values most appropriate to our model. Current medical literature and sensitivity analysis (see subsection 3.3) allow us to conclude that the corresponding interactions are of importance in the description of immune response.

**3. Non-dimensionalization, steady state and sensitivity analysis**

*3.1. Scaling*

For convenience, let us introduce dimensionless variables and parameters as follows: $T' = T/T_0$, $L' = L/L_0$, $I'_2 = I_2/I_{20}$, and $t' = t/\tau$ where $T_0 = 2.6 \times 10^6$ cells, $L_0 = 10^6$ cells, $I_{20} = 2 \times 10^7$ cells, and $\tau = \beta_L^{-1}$. The time-scale factor $\tau$ is chosen on the basis that the mean lifespan of CTL is about three days and the similar time is needed for the proliferation of CTL and IL-2 production [6,11].

Dropping primes for notational clarity, one finally obtains the following scaled model

$$\dot{T} = -h_1 T \ln \frac{h_2 T}{h_1} - h_3 TL, \tag{11}$$

$$\dot{L} = h_4 + h_5 LI_2 - L, \tag{12}$$

$$\dot{I}_2 = \frac{h_6 T}{T + h_9} - h_7 LI_2 - h_8 TI_2, \tag{13}$$

where $h_1 = \alpha_T/\beta_L$, $h_2 = \beta_T T_0/\beta_L$, $h_3 = \gamma'_L L_0/\beta_L$, $h_4 = V_L/\beta_L L_0$, $h_5 = \alpha_L I_{20}/\beta_L$, $h_6 = \alpha_{I_2}/\beta_L I_{20}$, $h_7 = \tilde{\alpha}_L L_0/\beta_L$, $h_8 = \gamma_T T_0/\beta_L$ and $h_9 = K_T/T_0$.

*3.2. Steady states analysis*

Let us perform a steady state analysis of the system (11)—(13) by using isoclines. We consider the phase plane $TL$ to reflect interactions between two main populations: tumor cells and CTL. In this case, the equations for main isoclines read

$$(h_4 - L)(T + h_9)(h_7 L + h_8 T) + h_5 h_6 TL = 0 , \qquad (14)$$

$$T = 0 , \quad L = -\frac{h_1}{h_3} \ln \frac{h_2 T}{h_1} . \qquad (15)$$

The fixed points are situated at the intersections of isoclines (14) and (15). Our analysis shows that the system (11)—(13) has an unstable point $(0, h_4, 0)$ for any choice of parameters. This point lies at the intersection of isoclines (14) and $T = 0$.

We consider $\alpha_{I_2}$ as a varying parameter to present possible model outcomes. In fact, $\alpha_{I_2}$ features the antigen presentation. Indeed, it is proportional to $\gamma_H$ which characterizes the probability of interaction between mAPC and HTLP. In turn, this probability depends on the expression of AG-MHC-II complexes on the surface of APC. The antigen presentation by APC is considered as one of important factors in the immune response to tumor. Tumor cells develop a number of mechanisms to escape recognition and elimination by immune system. One of them is the loss or down-regulation of MHC class I and II molecules presenting AG on tumor cells. This mechanism prevents lymphocytes from recognizing tumor cells [38]. If tumor cells do not possess antigens of MHC-II, an activation of helper T cells depends on the processing of tumor antigens by APC.

A bifurcation diagram for the dimensionless parameter $h_6$ is presented in Figure 2 where the function $h_6(T)$ is obtained by substitution of $L$ from (15) into (14). As is seen, there are three bifurcation points. Therefore one can distinguish four main dynamical regimes. For *a low antigen presentation* ($h_6 < h_{6min}$) the system (11)—(13) has two fixed points: a saddle point $(0, h_4, 0)$ and an improper node $(T_3, L_3, I_{23})$. This means that the population of tumor cells is able to escape from the immune response under IL-2 deficiency. The tumor grows and the immune system becomes suppressed. In the region $h_{6min} < h_6 < h_{6max}$ corresponding to *a medium antigen presentation* there appear two additional fixed points: a stable spiral $(T_1, L_1, I_{21})$ and an unstable saddle $(T_2, L_2, I_{22})$. Therefore different regimes can exist depending on the initial conditions. First, when the initial size of CTL population is sufficiently large the regression of tumor up to a small fixed size takes place (the dynamical equilibrium between tumor and immune system is reached). In this case, the tumor manifests itself via the excited immune system. Second regime appears when initial number of CTL is not large enough to drive the system at the dynamical equilibrium point $(T_1, L_1, I_{21})$, which is a stable spiral. Thus, the tumor grows to a highest possible size defined by conditions of restricted feeding. The dynamical equilibrium between the tumor and the immune system is reached at the fixed point $(T_3, L_3, I_{23})$ that is an improper node. In the case of *a high antigen presentation* ($h_6 > h_{6max}$) the fixed points $(T_2, L_2, I_{22})$ and $(T_3, L_3, I_{23})$ disappear. As a result, there are two fixed points: a saddle point $(0, h_4, 0)$ and a stable spiral $(T_1, L_1, I_{21})$. In this case, a decrease in tumor size is found when the equilibrium between the tumor and the immune system is established (dormant tumor). Finally, let us discuss the case of a high antigen presentation level ($h_6 > $ HB) when Hopf bifurcation occurs and stable spiral $(T_1, L_1, I_{21})$ becomes unstable spiral. Integral curves tend to stable limit cycle and, accordingly, we observe oscillations in small tumor size, number of CTL and the concentration of IL-2. This means that the immune system is able to prevent tumor from uncontrolled growing. This also corresponds to the dormant tumor.

*3.3. Sensitivity analysis*

The sensitivity analysis has been carried out to test which components of the model (8)—(10) contribute most significantly to tumor dynamics. We altered each parameter (taken separately) from its estimated value (Table 1, M1) by 1% and calculated the change in the tumor size after 30 days. The results are shown in Figure 3. As is seen, the system is most sensitive to the tumor growth rate $\alpha_T$ and the CTL death rate $\beta_L$.

We found lesser (yet remarkable) sensitivity to the following parameters: the rate of tumor cells inactivation by CTL $\gamma'_L$, the CTL proliferation rate $\alpha_L$, the antigen presentation $\alpha_{I_2}$ as well as the rate of inactivation of the IL-2 molecules by prostaglandins $\gamma_T$. The system is of little sensitivity to the consumption of IL-2 $\tilde{\alpha}_L$ and the half-saturation constant $K_T$. What is important for our consideration, the parameters $\gamma'_L$ and $\alpha_{I_2}$ belong to the second group. This means that even a small variation of either the antigen expression on tumor cells or the antigen presentation by APC will markedly affect tumor dynamics. Based on both bifurcation and sensitivity analysis we will associate the region I in Figure 2 with a weak immune response, and the region II with the strong immune response. The region III is associated with the case of dormant tumor when the immune system is able to handle the tumor size.

In conclusion, it is interesting to examine how alterations of either $\alpha_T$ or $\gamma'_L$ affect the model regimes. Let us introduce a variable $\tilde{T}$

$$\tilde{T} = \frac{h_1}{h_2}\exp(-\frac{h_3 h_4}{h_1}), \tag{16}$$

which is a zero of the function $h_6(T)$ (see Figure 2). The bifurcation diagram for dimensionless parameters $h_6$ versus $h_1$ is shown in Figure 4(a). As is seen, both $h_{6min}$ and $h_{6max}$ increase with $h_1$. For small rate of tumor growth, the region II diminishes and $\tilde{T}$ decreases in (16). In this case, the region II becomes inessential and the dynamical behavior is determined by the regions I and III. The final tumor size in the region I becomes small in comparison with the case of rapidly growing tumor. Besides, in the region III HB increases with $h_1$ (see Figure 4(a)). This means that slowly growing tumors are not able to evade even weak immune supervision. In the case of high rate of tumor growth the region II markedly extends and $\tilde{T}$ increases. Therefore, a high antigen presentation is required to reach the region III corresponding to dormant tumor and the possibility of tumor remission decreases with increasing tumor growth rate. In other words, the rate of tumor growth can give warning of malignance.

The next important characteristic determining the outcome of the disease is the expression of AG-MHC-I complexes on the surface of tumor cells. In our consideration, a level of this expression is characterized by the parameter $\gamma'_L$. Figure 4(b) shows the bifurcation diagram for $h_6$ versus $h_3$. As is seen, with $h_3$ increasing the region II vanishes and $\tilde{T}$ descends in (16). This means that the immune system is able to handle cancer. For small antigen expression, the strength of the immune response depends on the level of antigen presentation ($h_6$). Therefore, for tumors with poor immunogenicity (low antigen expression) a high antigen presentation on APC can be responsible for the strong immune response.

**4. Immunocorrective effects of radiofrequency electromagnetic waves**

In this section we discuss a possible way to take into consideration the influence of low-intensity electromagnetic microwaves within our model. Since the main effects have a complex nonlinear dependence on frequency, intensity, and other characteristics of EMR we suggest using a phenomenological approach. To justify our consideration let us present an overview of some important biological and physical aspects.

Above all, we would like to stress that our consideration is restricted to the frequency range 8—18 GHz and a low incident power ~1 $\mu$W/cm$^2$ because namely these characteristics of EMR were explored in recent experiments by Glushkova et al. [17]. Two important experimental findings should be mentioned. First, both the concentration of IL-2 in the serum of tumor-bearing mice and the production of this cytokine were found to be normalized after exposure to microwaves. Second, the yield of heat shock proteins-72 (HSP-72) by spleencytes was observed in both healthy and tumor bearing mice exposed to radiation. The last finding is rather surprising and could indicate the presence of cellular stress response under the exposure. As is known, HSP play a role of "molecular chaperones" binding to and stabilizing partially unfolded proteins, thus providing the cell with protection. However, our estimation of the specific absorption rate (SAR) by using the empirical model by Durney et al. [12] gives ~0.5 mW/kg for mouse. In experiments [17] mice were exposed to microwaves daily during 20 days. The duration of the exposure was 1.5 hour. It is easy to estimate that during 1.5 hours only 2.7 J/kg of electromagnetic energy is absorbed. Therefore, the intensity level used in [17] is not sufficient for occurring conformation changes. In this case, the question arises: how to explain the appearance of HSP? Unfortunately, this is an open problem yet. Nevertheless, some existing ideas allow us to suggest the following scenario.

In accordance with a hypothesis of the resonant absorption the electromagnetic energy in microwave (RF) range is absorbed mainly by aqueous environment. Therefore, the observed HSP production could be caused by free radicals in water (see, e.g., [20]). According to [4,47], free radicals may be produced from water ($H_2O$) by any process that moves clusters of water relative to each other, for instance, the mechanical vibration

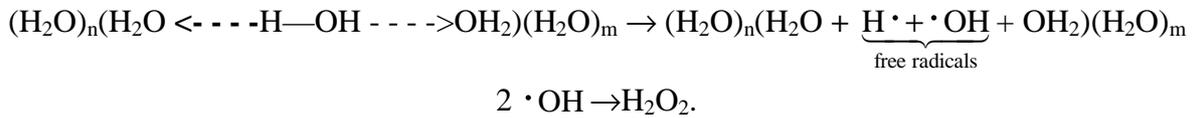

$$(H_2O)_n(H_2O \mathrel{<\text{- - -}} H\text{—}OH \mathrel{\text{- - -}>} OH_2)(H_2O)_m \rightarrow (H_2O)_n(H_2O + \underbrace{H\cdot + \cdot OH}_{\text{free radicals}} + OH_2)(H_2O)_m$$

$$2\cdot OH \rightarrow H_2O_2.$$

In the case of low-intensive EMR, small mechanical vibrations of water clusters may result from non-radiating transitions of excited molecules. It should be stressed that at low incident power of EMR very low concentrations of free radicals will be formed. This is very important for getting the therapeutic effect because the perturbations in concentrations of free radicals should not exceed physiological levels. In this case, mechanisms of natural antioxidant defense are able to reduce oxidative stress. For example, melatonin is found to mediate the inactivation of free radicals by stimulating some important antioxidative enzymes [36]. Besides, melatonin is able to activate helper T lymphocytes thereby increasing the production of IL-2 and IFN-γ [15]. This could explain the experimentally observed recovery of IL-2 production. There is also a different possible mechanism of antioxidant defense when free radicals activate such nucleus transcription factors as NFAT and NFκB (see [45] and the references therein). Indeed, NFκB and NFAT induce the expression of the antioxidant genes [20,45]. It has been recently observed in experiment that the production of NFκB actually increases as a result of exposure to weak RF EMR [23]. Notice that NFAT and NFκB are transcriptional regulators of the IL-2 gene [25,39]. Therefore, additionally to the antigen stimulation, these factors can be also activated by EMR-induced free radicals thereby enhancing the production of both IL-2 and very likely IFN-γ.

Let us revert to the model. In order to reflect the influence of EMR we assume to vary two basic model parameters $\gamma'_L$ and $\alpha_{I_2}$. Let us remind that $\gamma'_L$ represents the destruction rate of tumor cells by CTL. With growing production of IFN-γ the expression of molecules MHC class I and II on tumor cells increases thus enhancing their recognition by CTL [35]. In addition, HSP-72 also mediate up-regulation of AG- MHC-I complexes on surface of tumor cells [49]. Therefore, the parameter $\gamma'_L$ should be increased for taking into account the radiation. The parameter $\alpha_{I_2}$ characterizes the antigen presentation. Notice that for big tumor sizes $\alpha_{I_2}$ determines the rate of the IL-2 production that enhanced by the melatonin. Therefore $\alpha_{I_2}$ also should be increased. We assume that these parameters remain time-independent and merely increase to the new constant values $\gamma'_{L\exp}$ and $\alpha_{I_2\exp}$. In other words, we suggest that an influence of EMR is effective during all the time between exposures. Unfortunately, it is impossible to extract the values of $\gamma'_{L\exp}$ and $\alpha_{I_2\exp}$ from existing experiments. Therefore, we will study the role of these parameters by taking into account the fact that the influence of low-intensity EMR is weak. In this case, we use trial values for $\gamma'_{L\exp}$ and $\alpha_{I_2\exp}$ assuming that $\gamma'_L$ and $\alpha_{I_2}$ are only slightly increased under exposure (by 2% and 4%, respectively, see Table 1). As an additional criterion, the interval of variability of these parameters should be chosen in such a way to prevent the system from passing to the region III where the regime of dormant tumor is realized (see Figure 4(b)).

We present numerical results for two parameter sets M1 and M2 (see Table 1) to illustrate the body specific effects of electromagnetic radiation. Figure 5 shows bifurcation diagrams for both M1 and M2. As is seen, in both cases the system is located in the region of the strong immune response. Hence the outcome of disease depends on the initial conditions. We assume the same initial numbers of tumor cells and CTL whereas the initial concentration of IL-2 for M1 is taken to be higher than for M2. In this case, the remission of tumor for M1 and progressive growth for M2 are found (see Figure 6 and 7). As is seen from Figure 6, without exposure the dynamical curves for M1 have a character of dumping oscillations. The tumor decreases to a small size corresponding to the stable spiral. Although the tumor growth is handled by the immune system, for the first 20 days the tumor size is high enough (Figure 6(a)). As a result, the IL-2 concentration is smaller than its initial value during this period (Figure 6(c)). At the same time the population of CTL increases (Figure 6(b)). The results show that tumor cells stimulate immune response. This qualitatively agrees with the experimental results [17] where both the decrease of the IL-2 concentration and the increase of the number of CTL were observed in 20 days of tumor growth.

Figure 6(a) shows that after exposure to weak RF electromagnetic waves during 20 days the tumor size becomes smaller than in the case without exposure. The concentration of IL-2 markedly increases and reaches the initial value on 20th day (Figure 6(c)). Accordingly, the population of CTL also grows up to a larger value in comparison with the case without exposure (Figure 6(b)). Thus, our results show that the concentration of IL-2 is restored as a result of exposure to EMR, which also qualitatively agrees with the experimental observations [17]. It should be mentioned that there are some differences between predictions of our model and the experiment. For example, in experiment a decrease of the CTL population in comparison with unexposed mice was found after 20 days of irradiation instead of the increase in our model. It may be that the production of HSP blocking the proliferation is responsible for this observation. The dynamics of HSP is not explicitly taken into account in our model.

In the case of M2, without exposure the tumor grows up to the maximum possible value (Figure 7(a)). At the same time, the population of CTL and the IL-2 concentration decrease (Figures 7(b) and 7(c)). Nevertheless, initially the tumor stimulates the immune response. Hence, the number of CTL on 20th day of tumor growth is higher than their initial value (Figure 7(b)). As is seen form Figure 7, after cessation of daily exposure to weak RF EMR during 20 days (when the parameters take their normal (initial) values) the dynamical curves tend to the stable spiral, and the tumor remission takes place. At the same time, the population of CTL and the concentration of IL-2 increase in comparison with unexposed case. Thus, the behavior of the IL-2 concentration for M2 also qualitatively agrees with experimental observations [17]. It is important that the influence of weak EMR leads to the change of dynamical regime from progressive growth to remission of tumor. This follows from the fact that the number of tumor cells and CTL as well as the IL-2 concentration fall into the basin of attraction of stable spiral after the cessation of exposure. Summarizing, our results show the pronounced immunocorrective effect of the weak RF EMR.

# 5 Conclusion

In this paper, we have formulated the mathematical model for the immune response to the malignant growth with the IL-2 taken into account. It is found that tumor growth rate and the level of antigen expression on tumor cells and APC are important factors determining the dynamics of disease. Four main dynamical regimes are revealed and shown on the bifurcation diagram for antigen presentation by APC. For a low antigen presentation the tumor is able to escape from the immune response. In the case of a medium antigen presentation there exist two regimens of disease depending on both the initial tumor size and the condition of immune system: (i) the regression to small tumor when the dynamical equilibrium is established and (ii) a progressive tumor growth to the highest possible size. For a high antigen presentation the decrease of the tumor size is found when the equilibrium between the tumor and the immune system is established. Additionally, the regime of oscillations in small tumor size, number of CTL and the concentration of IL-2 is observed due to the presence of stable limit cycle. It is important to note that the regime of full tumor regression as a result of the immune response alone is not admitted within our model. This fact is in agreement with clinical observations where spontaneous regression of tumors is not possible.

In order to illustrate the behavior of the system with the effects of weak RF EMR taken into account we have chosen two parameter sets so that the system is located in the region II of bifurcation diagram where the result of immune response depends on initial tumor size and the immune system condition. Namely in this region the system is most sensitive to perturbation of the model parameters. We have considered the influence of two model parameters characterizing both the rate of inactivation of tumor cells by cytotoxic T cells and the production of IL-2. Our results show the marked immunocorrective effect of weak RF EMR. In particular, an increase of the IL-2 concentration in comparison with unexposed case and enhancement of the immune response are found. Moreover, it may be expected that the RF EMR at low intensity is low-toxic. Indeed, we found only minor increase of the IL-2 concentration which does not exceed the norm. Nevertheless, the frequency range, intensity, and other EMR parameters as well as the regimen of exposure should be carefully estimated to avoid the harmful influence of EMR on the central nervous, cardiovascular and other systems of the body.

**References**

[1] J.A. Adam, and N. Bellomo, *A survey of Models for Tumor-Immune System Dynamics*, Birkhäuser, Boston, MA, 1996.

[2] B. Alberts, D. Bray, J. Lewis, M. Raff, K. Roberts, and J.D. Watson, *Molecular biology of the cell* (3rd edition), New York: Garland Publishing Inc., 1408pp, 1994

[3] J.C. Arciero, D.E. Kirschner, and T.L. Jackson, *A mathematical model of tumor-immune evasion and siRNA treatment*, Disc. Cont. Dyn. Syst.-B 4(1) (2004), pp. 39—58.

[4] H.J. Bakker and H.-K. Nienhuys, *Delocalization of protons in liquid water*, Science, 297 (2002), pp. 587-590.

[5] A. Casrouge, E. Beaudoing, S. Dalle, C. Pannetier, J. Kanellopoulos, and P. Kourilsky, *Size estimate of the ab TCR repertoire of naive mouse splenocytes*, The Journal of Immunology, 164 (2000), pp. 5782–5787.

[6] D.L. Chao, M.P. Davenport, S. Forrest, and A.S. Perelson, *A stochastic model of cytotoxic T cell responses*, Journal of Theoretical Biology, 228 (2004), 227—240.

[7] M. Chaplain and A. Matzavinos, *Mathematical modeling of spatio-temporal phenomena in tumor immunology*, Lect. Notes Math. 1872 (2006), pp. 131-183

[8] M.A.J. Chaplain, *Mathematical models in cancer research*. in: *The Cancer Handbook*, Nature pub-lishing group. Chapter 60, 2003, pp. 937—951.

[9] S.F. Cleary, L.M. Liu, and R.E. Merchant, *Lymphocyte proliferation induced by radio-frequency electromagnetic radiation under isothermal conditions*, Bioelectromagnetics, 11 (1990), pp. 47-56.

[10] R.J. De Boer, P. Hogeweg, F.J. Dullens, R.A. De Weger and W. Den Otter, *Macrophage T lymphocyte interactions in the anti-tumor immune response: a mathematical model*, J. Immunol. 134(4) (1985), pp. 2748—2758.

[11] R.J. De Boer, M. Oprera, R. Antia, K. Murali-Krishna, R. Ahmed, and A.S. Perelson, *Recruitment times, proliferation, and apoptosis rates during the CD8+ T cell Response to lymphocytic choriomeningitis virus*, Journal of Virology, 75(22) (2001), pp. 10663—10669.

[12] C.H. Durney, M.F. Iskander, H. Massoundi, and C.C. Johnson, *An empirical formuls for broadband SAR calculation of prolate spheroidal models of humans and animals*. IEEE Trans. Microwave theory techn., MTT-27(8) (1979), pp. 758—763.

[13] R. Garay and R. Lefever, *A kinetic approach to the immunology of cancer: stationary states properties of effector-target cell reactions*, J. Theor. Biol., 73 (1978), pp. 417—438.

[14] S. Garbelli, S. Mantovani, B. Palermo, and C. Giachino, *Melanocyte-specific, cytotoxic T cell responses in vitiligo: the effective variant of melanoma immunity*. Pigment Cell Res. 18 (2005), pp. 234—242.

[15] S. Garcia-Maurino, M.G. Gonzalez-Haba, J.R. Calvo, M. Rafii-El-Idrissi, V. Sanchez-Margalet, R. Goberna, and J.M. Guerrero, *Melatonin enhances IL-2, IL-6, and IFN-gamma production by human circulating CD4+ cells: a possible nuclear receptor-mediated mechanism involving T helper type 1 lymphocytes and monocytes*. The Journal of Immunology, 159(2) (1997), pp. 574-581.

[16] B.L. Gause, M. Sznol, W.C. Kopp, J.E. Janik, J.W. Smith II, R.G. Steis, W.J. Urba, W. Sharfman, R.G. Fenton, S.P. Creekmore, J. Holmlund, K.C. Conlon, L.A. VanderMolen, and D.L. Longo, *Phase I study of subcutaneously administered interleukine-2 in combination with interferon alfa-2a in patients with advanced cancer*, J. of Clin. Oncol. 14(8) (1996), pp. 2234—2241.

[17] O.V. Glushkova, E.G. Novoselova, O.A. Sinotova, and E.E. Fesenko, *Immunocorrecting effect of super-high frequency electromagnetic radiation in carcinogenesis in mice*. Biophysics, 48(2) (2003), pp. 264-271.

[18] I. Hara, H. Hotta, N. Sato, H. Eto, S. Arakava, and S. Kamidono, *Rejection of mouse renal cell carcinoma elicited by local secretion of interleukin-2*, J. Cancer Res. 87 (1996), pp. 724—729.

[19] S. Heegaard, M. Spang-Thomsen, and J.U. Prause, *Establishment and characterization of human uveal malignant melanoma xenografts in nude mice*. Melanoma Research, 13(3) (2003), pp. 247-251.
12

**Table caption**

Table 1. Parameter sets

| Parameter | Units | Description | Value M1 | Value M2 | Source |
|---|---|---|---|---|---|
| $\alpha_T$ | day$^{-1}$ | Tumor growth rate | 0.22 | | Fit to data [34] |
| $\beta_T$ | cell$^{-1}$day$^{-1}$ | $a_\xi/b_\xi$ is tumor carrying capacity | 8.4×10$^{-8}$ | | Fit to data [34] |
| $\gamma'_L$ | cell$^{-1}$ day$^{-1}$ | Rate of tumor cells inactivation by CTL | 4×10$^{-7}$ | 2.8×10$^{-7}$ | |
| $\gamma'_{L\exp}$ | | | 4.08×10$^{-7}$ | 2.86×10$^{-7}$ | |
| $V_L$ | cell day$^{-1}$ | Rate of steady inflow of CTL | 7.9×10$^4$ | | Estimated using [35], [36] |
| $\alpha_L$ | cell$^{-1}$ day$^{-1}$ | CTL proliferation rate induced by IL-2 | 9.9×10$^{-9}$ | 1.12×10$^{-8}$ | |
| $\beta_L$ | day$^{-1}$ | CTL death rate | 0.33 | | Estimated from [37] |
| $\alpha_{I_2}$ | unit day$^{-1}$ | Antigen presentation | 1.25×10$^7$ | | |
| $\alpha_{I_2\exp}$ | | | 1.3×10$^7$ | | |
| $\tilde{\alpha}_L$ | cell$^{-1}$ day$^{-1}$ | Rate of consumption of IL-2 by CTL | 6.6×10$^{-8}$ | | |
| $\gamma_T$ | cell$^{-1}$ day$^{-1}$ | Inactivation of IL-2 molecules by prostaglandines | 6.6×10$^{-7}$ | 5.5×10$^{-7}$ | |
| $K_T$ | cell | Half-saturation constant | 5.2×10$^4$ | 1×10$^5$ | |

**Figure Captions**

**Figure 1.** A scheme of the T cell mediated immune response.

**Figure 2.** The bifurcation diagram varying the antigen presentation ($h_6$). For $h_6 < h_{6min}$ there is only one steady state — improper node (region I). When $h_{6min} < h_6 < h_{6max}$ there are two stable steady states — improper node and spiral as well as an unstable (saddle) point (region II). For $h_6 > h_{6max}$ only one steady state, the stable spiral remains (region III). For $h_6 >$ HB the stable spiral passes to the stable limit cycle.

**Figure 3.** The sensitivity analysis for the parameter set M1 in Table1. The tumor size is more sensitive to tumor growth rate variable *a*, to CTL death rate *f*, to inactivation of tumor cells by CTL *c*, to antigen presentation *g*, to CTL proliferation variable *e* as well as to the rate of inactivation of the IL-2 molecules by prostaglandins *k*.

**Figure 4.** The bifurcation diagram $h_6$ vs $h_1$ (a). The bifurcation diagram $h_6$ vs $h_3$ and the variation of steady state regime under exposure to low-intensive RF EMR (b). Region I – weak immune response, region II – strong immune response, and region III – dormant tumor.

**Figure 5.** Bifurcation diagrams showing the steady state regimes for the model parameter sets M1 and M2.

**Figure 6.** Effects of low-intensive RF EMR: (a) tumor cells, (b) cytotoxic T cells, and (c) IL-2 vs time for the parameter set M1. The irradiation occurs during 20 days. Initial conditions: $2\times10^5$ tumor cells, $2.4\times10^5$ cytotoxic T lymphocytes, $3.6\times10^7$ IL-2 units.

**Figure 7.** Effects of low-intensive RF EMR: (a) tumor cells, (b) cytotoxic T cells, and (c) IL-2 vs time for the parameter set M2. The irradiation occurs during 20 days. Initial conditions: $2\times10^5$ tumor cells, $2.4\times10^5$ cytotoxic T lymphocytes, $2.4\times10^7$ IL-2 units.

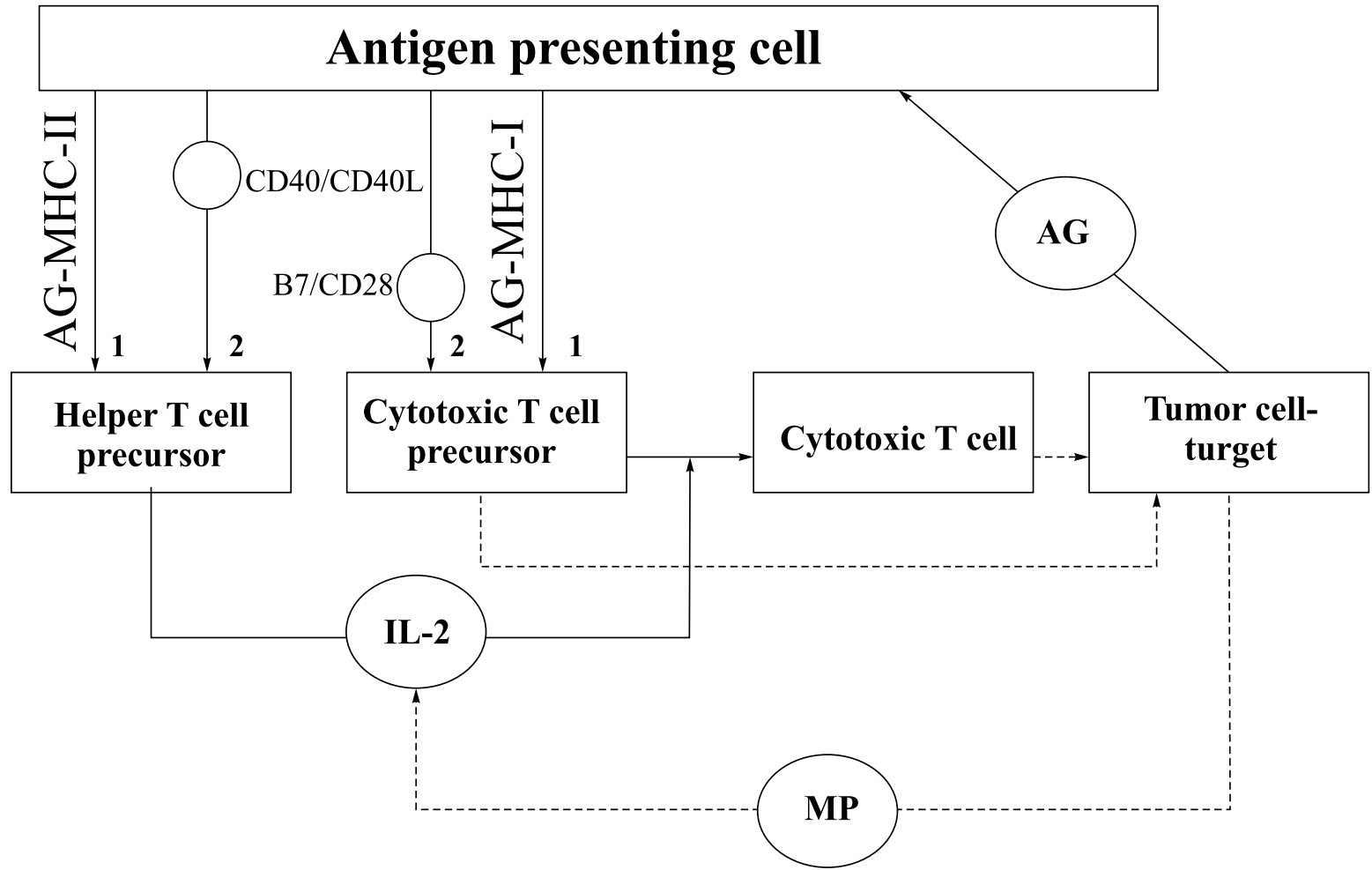

Figure 1

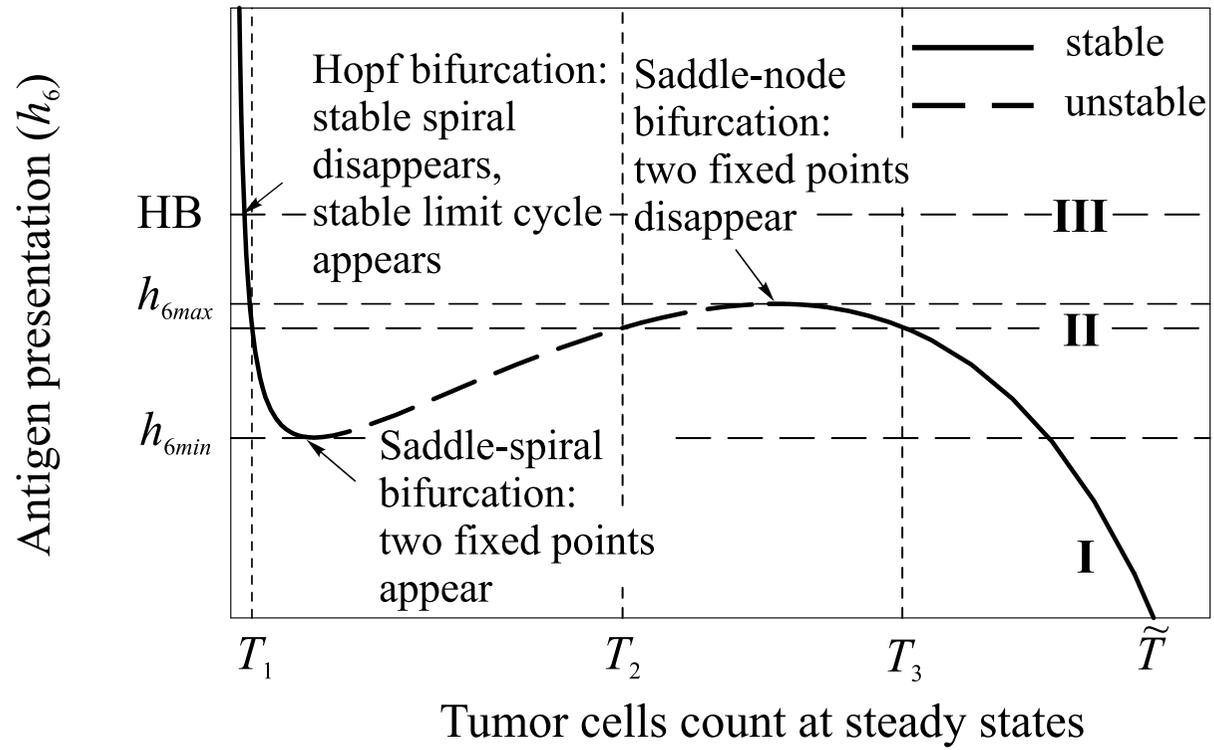

Figure 2

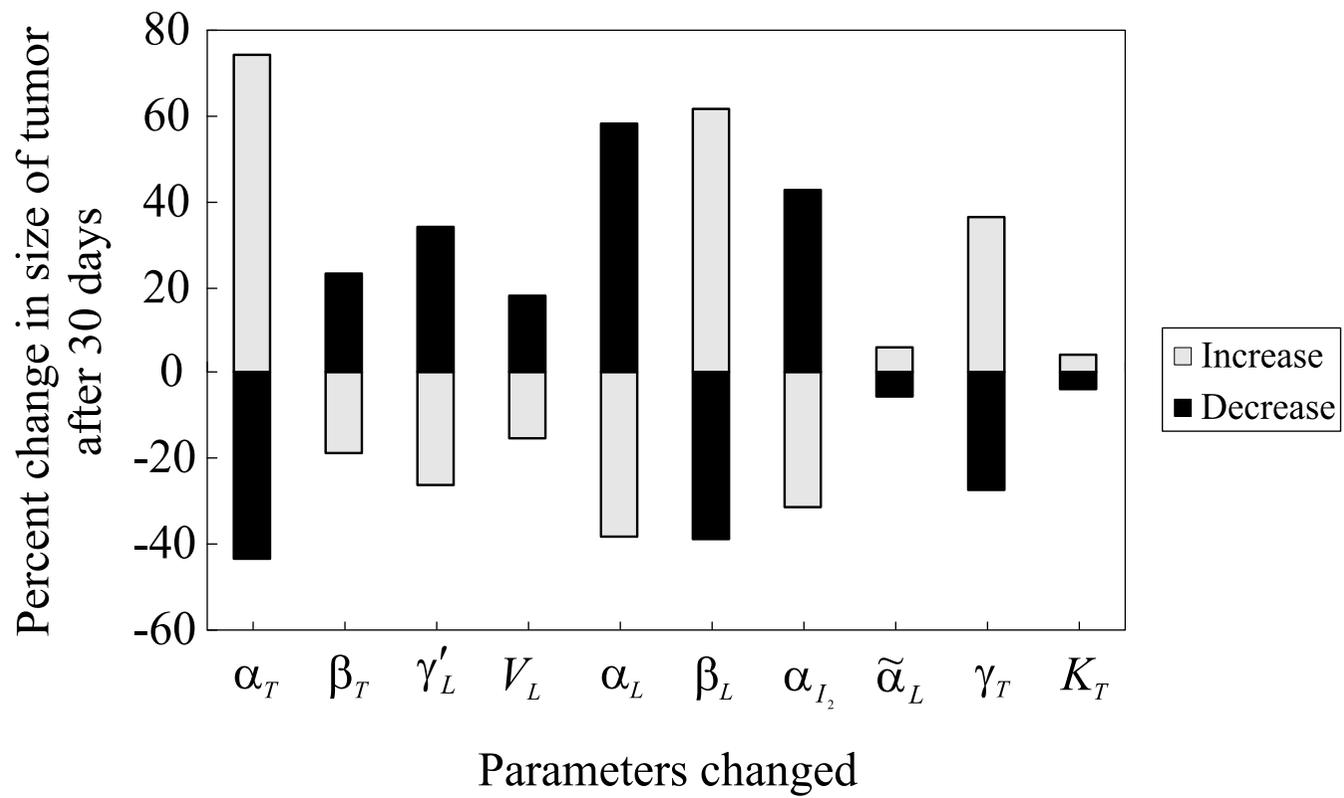

Figure 3

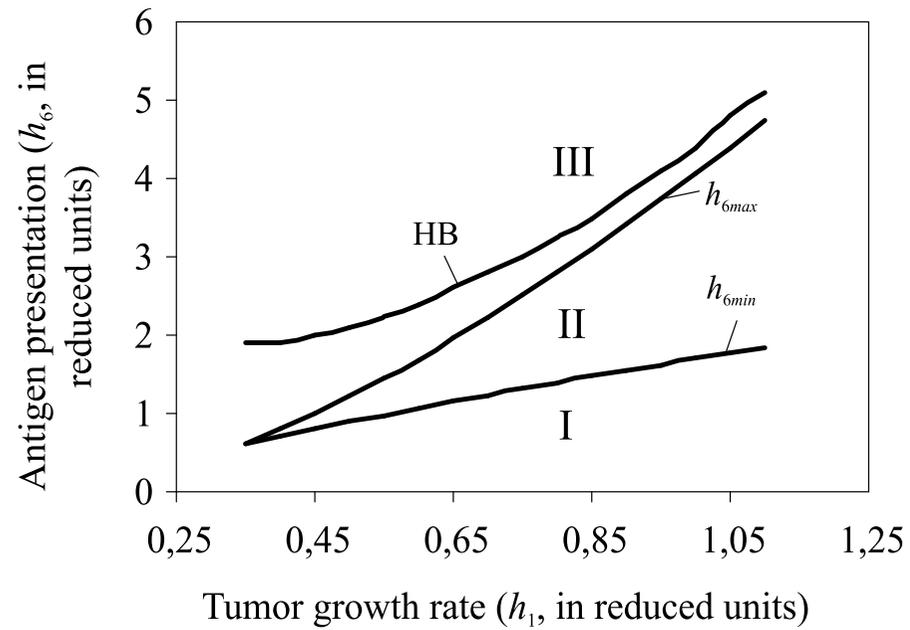 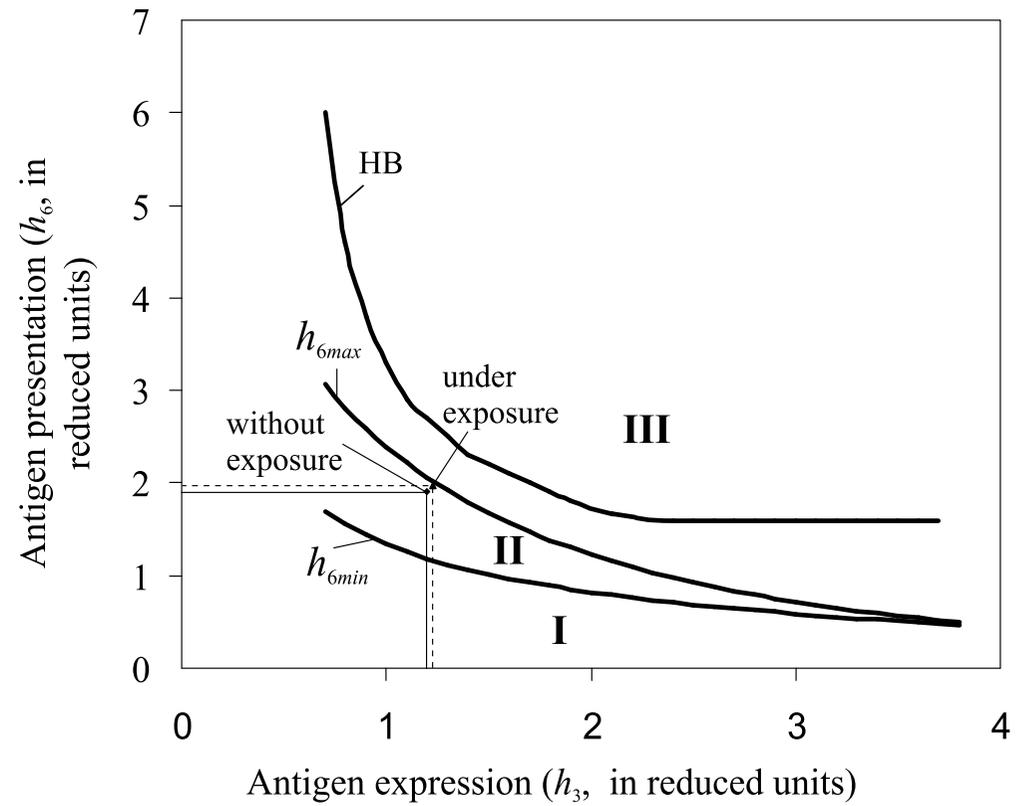

(a) (b)

Figure 4

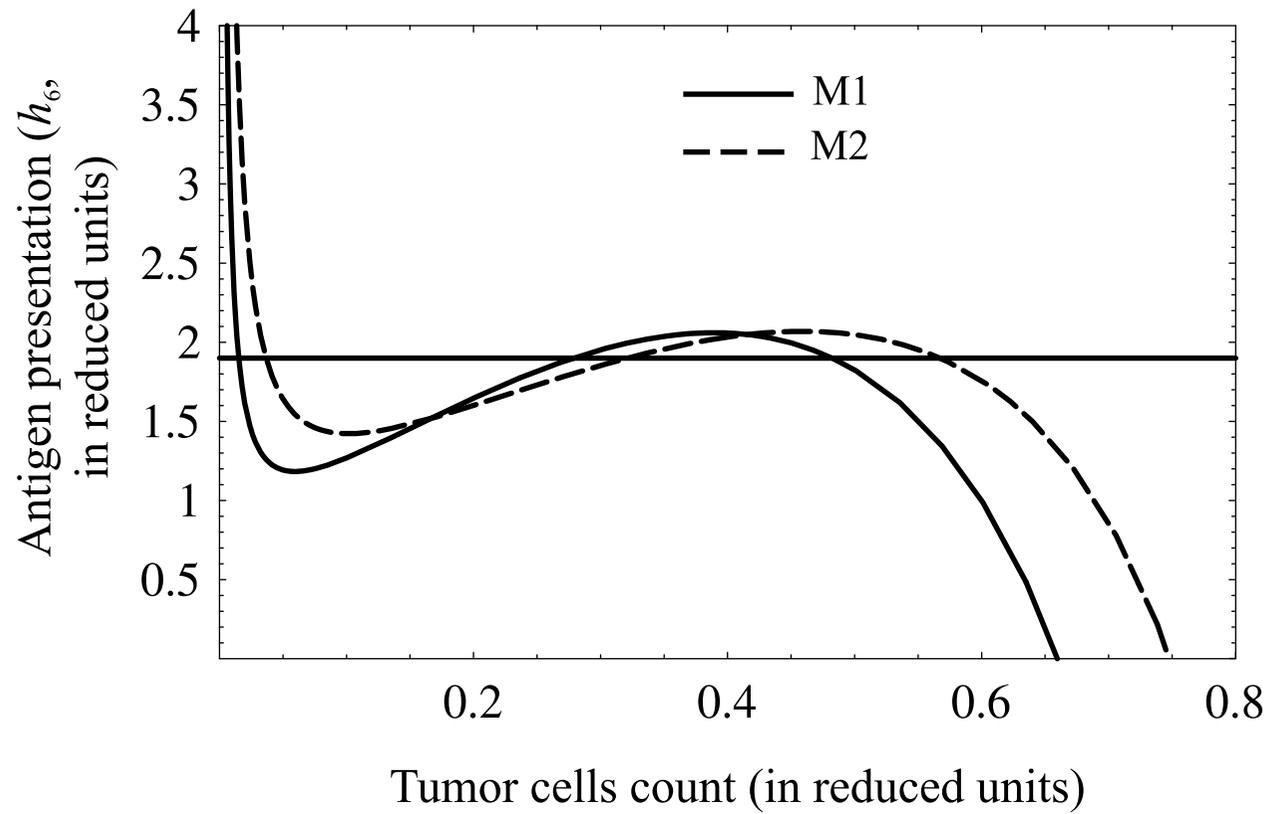

Figure 5

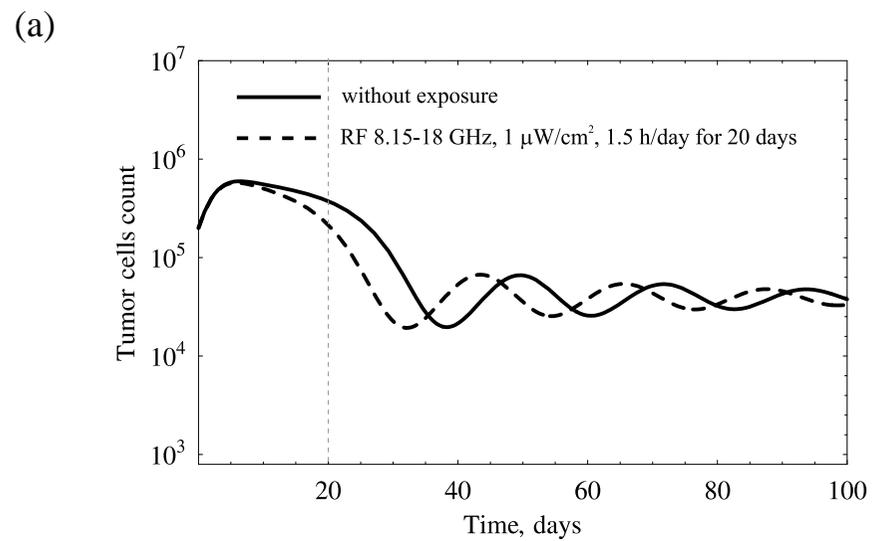
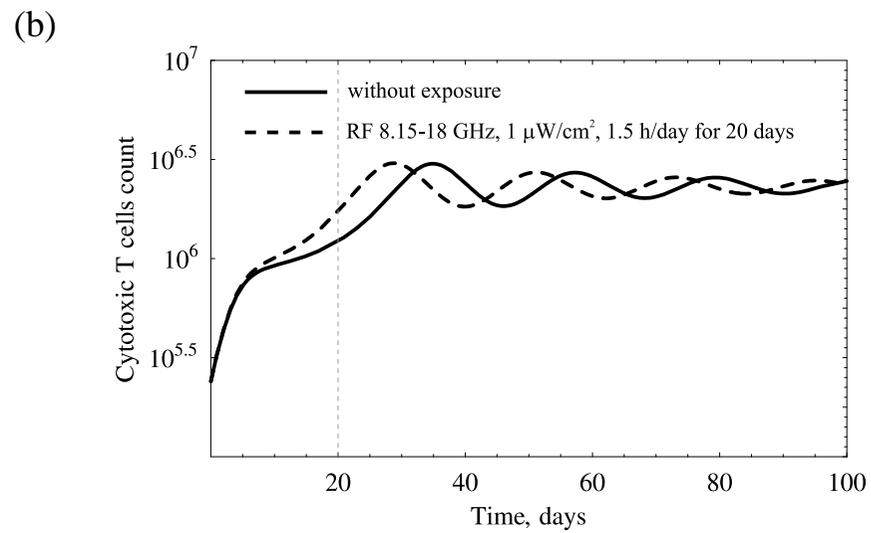
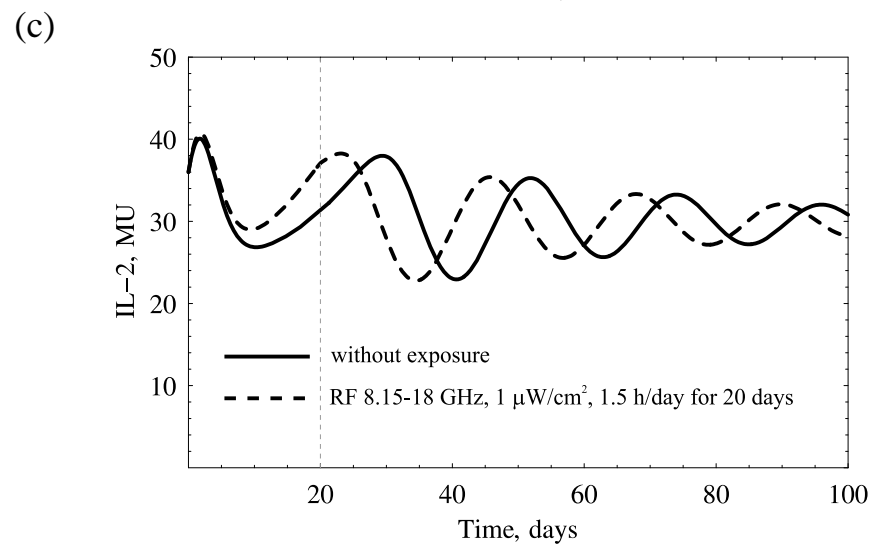

Figure 6

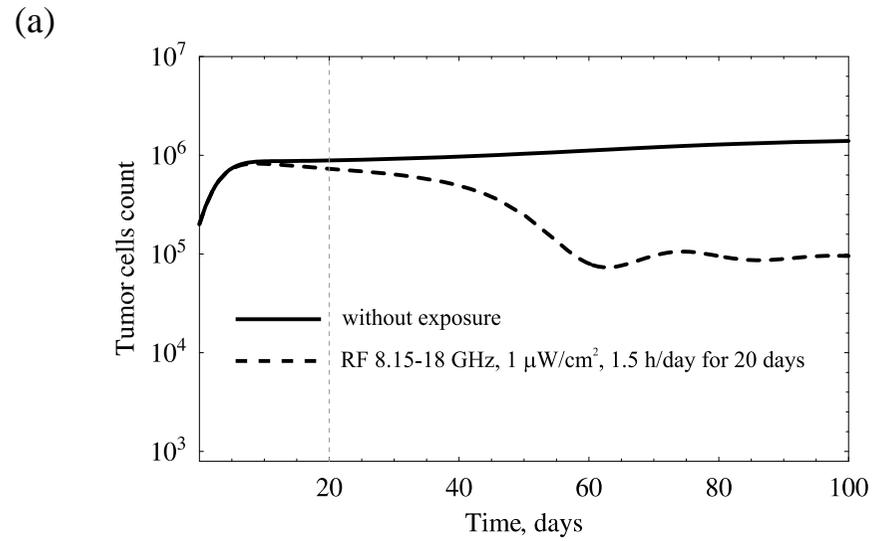
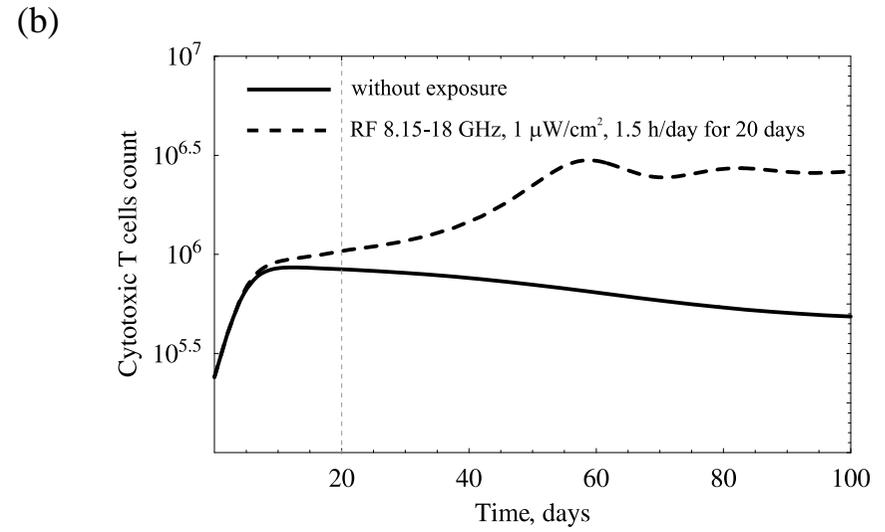
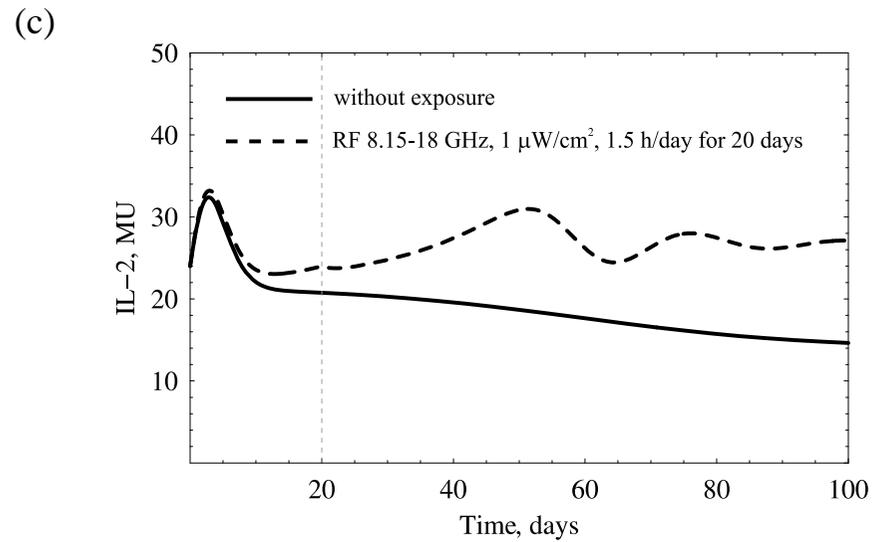

Figure 7